\documentclass[12pt]{article}

\usepackage[T1]{fontenc}
\usepackage[utf8]{inputenc}
\usepackage{amsmath}
\usepackage{amsfonts}
\usepackage{graphicx}
\usepackage[super,sort&compress,comma]{natbib}
\usepackage{booktabs}
\usepackage{a4wide}
\usepackage[doublespacing]{setspace}
\usepackage{hyperref}

\usepackage[cm]{sfmath}

\begin{document}

\begin{center}

\LARGE \textbf{A variational approach to nucleation simulation} \\

\vspace{0.5cm}

\large Pablo M. Piaggi$^{1,3}$, Omar Valsson$^{2,3}$, and Michele Parrinello$^{2,3,*}$ \\

\vspace{0.5cm}

{\normalsize
$^{1}$ Theory  and  Simulation  of  Materials  (THEOS) and  National  Centre  for Computational  Design  and  Discovery  of  Novel  Materials  (MARVEL), {\'E}cole Polytechnique F{\'e}d{\'e}rale de Lausanne,  CH-1015  Lausanne,  Switzerland
$^{2}$Department of Chemistry and Applied Biosciences, ETH Zurich, c/o USI Campus, Via Giuseppe Buffi 13, CH-6900, Lugano, Switzerland \\
$^{3}$Facoltà di Informatica, Instituto di Scienze Computazionali, and National Center for Computational Design and Discovery of Novel Materials (MARVEL), Università della Svizzera italiana (USI), Via Giuseppe Buffi 13, CH-6900, Lugano, Switzerland  \\ 
}
\normalsize $^*$E-mail: parrinello@phys.chem.ethz.ch

\end{center}

%%%%%%%%%%%%%%%%%%%%%%%%%%%%%%%%%%%%%%%%%%%%%%%%%%%%%%%%%%%%%%%%%%%%%
%% Abstract
%%%%%%%%%%%%%%%%%%%%%%%%%%%%%%%%%%%%%%%%%%%%%%%%%%%%%%%%%%%%%%%%%%%%%

\begin{abstract}
We study by computer simulation the nucleation of a supersaturated Lennard-Jones vapor into the liquid phase.
The large free energy barriers to transition make the time scale of this process impossible to study by ordinary molecular dynamics simulations.
Therefore we use a recently developed enhanced sampling method [Valsson and Parrinello, Phys.\ Rev.\ Lett.\ 113, 090601 (2014)] based on the variational determination of a bias potential.
We differ from previous applications of this method in that the bias is constructed on the basis of the physical model provided by the classical theory of nucleation.
We examine the technical problems associated with this approach.
Our results are very satisfactory and will pave the way for calculating the nucleation rates in many systems.
\end{abstract}


\section{Introduction}

Nucleation is a process that plays a prominent role in chemistry, engineering and materials science.
Among many others, it finds applications in the pharmaceutical industry where crystal shape and structure dramatically affect the drugs potency \cite{Chow08}.
Moreover, nucleation is not only technologically relevant but also scientifically interesting since it is a paradigmatic example of a self-assembly process \cite{Jacobs16}.
There is thus a great interest in understanding, and eventually controlling, the way in which a new phase emerges from the parent one \cite{Vekilov10,Davey13}.
In spite of the importance of nucleation, the small scales involved represent a formidable hurdle to experimental studies and thus the details of the process are not easy to unveil.
In this regard, molecular simulation and theory could pave the way to a better understanding of the early stages of this process.

One of the simplest examples of homogeneous nucleation is the condensation of a supersaturated vapor.
Being a fairly well understood phenomenon, it provides a useful scenario to test new simulation methods.
A vapor at a constant temperature and a pressure higher than the vapor pressure is in metastable equilibrium with respect to the liquid phase \cite{KashchievBook} and thus it will make a transition to the liquid phase in order to minimize its free energy.
However, one of the characteristics of first order phase transitions is the ability to remain in the metastable state due to the existence of a free energy barrier.
In this metastable state the system experiences density fluctuations that can be described as the fleeting appearence of small clusters of the new phase.
Occasionally the fluctuations are so large that a critical cluster is formed and the whole system condensates.
In order for this to occur, the system has to overcome a large energy barrier.
This makes nucleation a rare event that cannot be sampled in ordinary simulations.
Many solutions to this problem have been proposed \cite{Torrie77,Laio02,Bolhuis02} and applications to liquid-vapor nucleation have also been reported \cite{Chen01,Kusaka99,Merikanto04}.
The closest in spirit to our approach is the work of ten Wolde and Frenkel \cite{tenWolde98}.
In this work the free energy landscape of a Lennard-Jones fluid was explored via Monte Carlo (MC) simulations that used umbrella sampling\cite{Torrie77} to overcome kinetic barriers.

Our paper has a strong methodological connotation since we aim at applying to this well-studied problem the newly developed variationally enhanced sampling (VES) method\cite{Valsson14} in which the bias is determined via a variational procedure based on a suitably defined functional.
A few applications of the method have already been presented in the literature \cite{Valsson14,McCarty15,Valsson15,Shaffer16,McCarty16}.
The default way of solving the variational problem is to expand the bias potential in a basis set and use the expansion coefficients as variational parameters.
We differ from this approach in that, rather than following this procedure, we use a physically motivated expression derived from classical nucleation theory (CNT).
This expression contains two empirical parameters, the supersaturation and the effective surface energy.
We shall optimize the functional with respect to these two parameters.
This will allow us to understand better the properties of the functional and lay the foundations for future work in which we plan to use a variant of the variational method \cite{McCarty15,Tiwary13} designed to calculate nucleation rates.

%%%%%%%%%%%%%%%%%%%%%%%%%%%%%%%%%%%%%%%%%%%%%%%%%%%%%%%%%%%%%%%%%%%%%%%%%%%%%%%%
\section{Classical nucleation theory}
%%%%%%%%%%%%%%%%%%%%%%%%%%%%%%%%%%%%%%%%%%%%%%%%%%%%%%%%%%%%%%%%%%%%%%%%%%%%%%%%

\label{sec:ClassicalNucleationTheory}

The textbook way of describing nucleation phenomena is CNT.
In CNT the cost of forming a cluster of the new phase (in our case the liquid one) can be expressed as:
\begin{equation}
  \Delta F^{CNT}(n) = -\Delta\mu \: n + \sigma \: n^{2/3},
  \label{eq:reversibleWorkCNT}
\end{equation}
where $n$ is the number of atoms in the cluster, $\Delta\mu$ is the difference in chemical potential between the two phases (supersaturation), and $\sigma$ is an effective interfacial energy.
The first term represents the energy gain in going into the new more stable phase whereas the second term expresses the energetic cost of forming an interface between the liquid and the vapor.
Since on average liquid clusters are spherical, $\sigma$ can be related to the surface free energy $\gamma$ by $\sigma= (36 \pi)^{1/3} \rho^{-2/3} \gamma $ where $\rho$ is the density of the new phase.
The supersaturation $\Delta\mu$ is often expressed as a dimensionless quantity called supersaturation ratio ($S$) and defined through $\Delta\mu=k_B T \log S$. 
The expression in equation \eqref{eq:reversibleWorkCNT} differs from the one normally found in textbooks in that the latter uses the drop radius $R$ as a measure of its size.
If one makes the assumption that the drops have a spherical shape, the relation between $n$ and $R$ is $n=\frac{4}{3} \pi R^3 \rho$.
Substituting this relation in equation \eqref{eq:reversibleWorkCNT}, the standard expression $\Delta F^{CNT}(R) = -\frac{4}{3} \pi  R^3 \: \rho \: \Delta\mu+  4 \pi R^2 \: \gamma$ is recovered.

%%%%%%%%%%%%%%%%%%%%%%%%%%%%%%%%%%%%%%%%%%%%%%%%%%%%%%%%%%%%%%%%%%%%%%%%%%%%%%%%
\section{Methods}
%%%%%%%%%%%%%%%%%%%%%%%%%%%%%%%%%%%%%%%%%%%%%%%%%%%%%%%%%%%%%%%%%%%%%%%%%%%%%%%%

The presence of high barriers often hinders exhaustive sampling in molecular simulations.
Enhanced sampling methods aim at solving this problem.
Many rely on the introduction of a bias potential $V$ which is a function of a small number of collective coordinates $\textbf{s}$.
Among these, it is worth noting the historically important umbrella sampling \cite{Torrie77}, also used by ten Wolde and Frenkel \cite{tenWolde98}, and metadynamics which has proved to be succesful in a variety of fields \cite{Laio02,Valsson16review}.
In the next section we describe the theoretical underpinnings of the recently introduced VES \cite{Valsson14}.

\subsection{Variationally enhanced sampling}

\label{subsec:VES}
As in umbrella sampling and many other enhanced sampling methods, we project the high-dimensional $\mathbf{R}$ space of the $N$ particle system into a much smaller and smoother $d$-dimensional space by introducing the set of collective variables $\mathbf{s}(\mathbf{R})=(s_1(\mathbf{R}),s_2(\mathbf{R}),...,s_d(\mathbf{R}))$ that give a coarse-grained description of the system.
The free energy surface (FES) associated to the CV set $\mathbf{s}$ is defined as:
\begin{equation}
F (\mathbf{s}) = -\frac{1}{\beta} \log \int d\mathbf{R} \: \delta\left(\mathbf{s} - \mathbf{s}(\mathbf{R})\right) e^{-\beta U(\mathbf{R})},
\label{eq:freeEnergyFunctionOfS}
\end{equation}
where we have dropped an immaterial constant as we shall also do in the following.

In ref.\ \citenum{Valsson14} it was shown how to construct a bias potential $V(\textbf{s})$ that acts on the CVs via the optimization of the following functional:
\begin{align}
\label{omega1}
\Omega [V] & =
\frac{1}{\beta} \log
\frac
{\int d\mathbf{s} \, e^{-\beta \left[ F(\mathbf{s}) + V(\mathbf{s})\right]}}
{\int d\mathbf{s} \, e^{-\beta F(\mathbf{s})}}
+
\int d\mathbf{s} \, p(\mathbf{s}) V(\mathbf{s}),
\end{align}
where $p(\mathbf{s})$ is a chosen target probability distribution.
This functional is convex and it is made stationary by the bias potential:
\begin{equation}
\label{optimal_bias}
V(\mathbf{s}) = -F(\mathbf{s})-{\frac {1}{\beta}} \log {p(\mathbf{s})}.
\end{equation}
It follows that once the functional is minimized, the probability distribution of $\mathbf{s}$ in the biased ensemble $p_V(\mathbf{s})$ is equal to the target distribution $p(\mathbf{s})$:
\begin{equation}
\label{p_of_s_biased_ensemble}
p_V(\mathbf{s})=
\frac{e^{-\beta [F(\mathbf{s})+V(\mathbf{s})] }}{\int d\mathbf{s} \: e^{ -\beta [F(\mathbf{s})+V(\mathbf{s})] }}
=p(\mathbf{s}).
\end{equation}

The simplest choice for $p(\mathbf{s})$ is to consider the uniform target distribution $p(\mathbf{s})=1/\Omega_s$, where $\Omega_s=\int d\mathbf{s}$ is the volume of CV space. 
In this case $F(\mathbf{s})=-V(\mathbf{s})$ as in standard metadynamics.
If instead one takes $p(\mathbf{s}) \propto p_0(\mathbf{s})^{1/\gamma}$ where $\gamma$ is greater than one and $p_0(\mathbf{s})$ is the equilibrium probability distribution of $\mathbf{s}$ in the unbiased ensemble \cite{Valsson15}, the well-tempered metadynamics distribution is recovered.
Other choices have been suggested \cite{McCarty15,Shaffer16} and here we shall also make use of the added flexibility offered by the freedom of choosing $p(\mathbf{s})$.
Once the bias has been determined, a standard reweighting procedure can be used to calculate statistical averages in the unbiased ensemble \cite{Torrie77}.
Details of the reweighting procedure are provided in the Supplementary Information (SI).

\subsection{Collective variable}

In order to use the CNT free energy expression to construct the bias, we need to define properly a CV that expresses in analytical and differentiable form the variable $n$ in equation \eqref{eq:reversibleWorkCNT} as a function of $\mathbf{R}$.
This requires first defining what is meant by liquid cluster.
For this, we follow the procedure suggested by ten Wolde and Frenkel\cite{tenWolde98}.
A pair of atoms is considered to belong to the same cluster if their distance is below an assigned radius $r_c$ and each of them has at least $n_c$ neighbors within $r_c$.
Once the clusters have been defined, in order to write a CV that corresponds to the one in CNT one would have had to sort the clusters by their size and consider also their multiplicity, a procedure that would have been expensive and cumbersome.

In their work based on MC, ten Wolde and Frenkel\cite{tenWolde98} decided to use as CV the largest cluster.
The idea behind this choice is that, as the system climbs the barrier, only the largest cluster survives.
Eventually the free energy is reweighted to obtain the cluster size distribution ($n$) as needed in the CNT expression.
In the present work, that is based on MD, the use of the largest cluster as CV would have caused problems in the calculation of the forces.
In fact, the flag of the largest cluster can change abruptly from one set of atoms to another.
Although this could have been remedied, we preferred to use as CV the total number of liquid-like atoms ($n_l$), a quantity that is easy to calculate.
In the SI we describe in detail the calculation of $n_l$.
As we shall see, for small systems where the probability of observing several clusters is negligible, this is a good choice and the resulting free energy as a function of this variable is similar to the CNT expression.
For larger systems this is no longer the case but still using a reweighting procedure the cluster size distribution can be obtained (see the SI for details).

\subsection{Using the CNT model for the bias potential}

Having decided how to represent $n$, albeit in an approximate form, we first take $p(s)$ to be uniform and write for the bias the functional form:
\begin{align}
V(s;\Delta\mu,\sigma)= & -\Delta F^{CNT}(s;\Delta\mu,\sigma) \nonumber \\
 = & - \left (-\Delta\mu \: s + \sigma \: s^{2/3} \right),
\label{eq:bias_model}
\end{align}
where $s=n_l$.
In equation \eqref{eq:bias_model} we have used the fact that if $p(s)$ is uniform, $V(s)=-F(s)$.
Expression \eqref{eq:bias_model} is then inserted into $\Omega[V]$ and the functional is minimized relative to $\Delta\mu$ and $\sigma$.
As already discussed in the introduction, this differs from the usual approach in which the bias $V(s)$ is expanded in an orthogonal basis set and the expansion coefficients are used as variational parameters.

\begin{figure}[t]
\centering \includegraphics[width=0.5\textwidth]{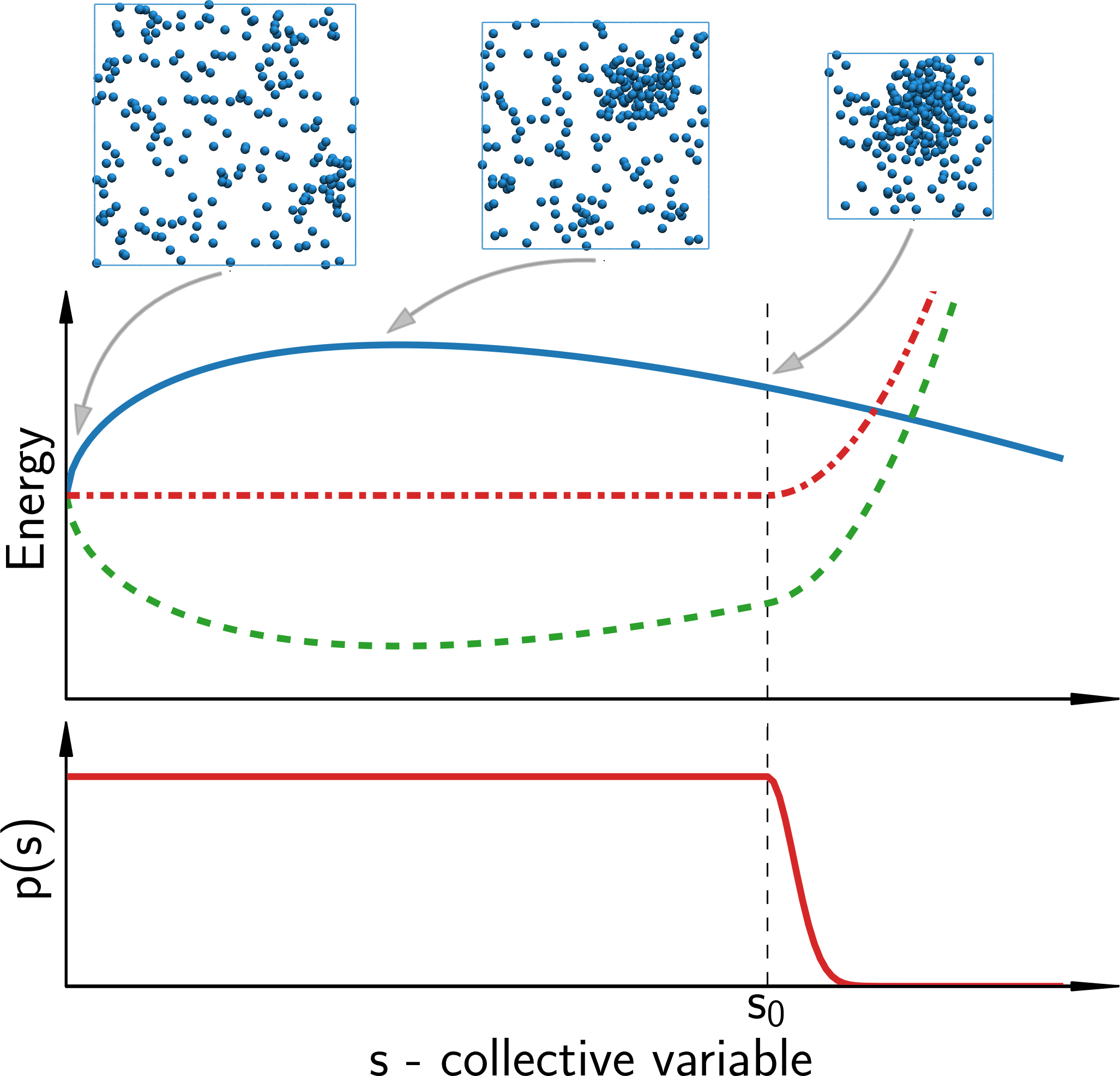}
\caption{\label{fig:methodIllustration}
Top plot) One dimensional nucleation free energy surface $F(s)$ (solid blue line), bias potential $V(s)$ (dashed green line), and effective free energy surface $F(s)+V(s)$ (dashed-dotted red line).
Bottom plot) Target distribution function $p(s)$ (solid red line).
The plots are here for illustrative purposes only and do not reflect the properties of specific physical systems.
}
\end{figure}

In principle at this stage we could have moved to describe the calculation.
However, before doing so, a practical issue needs to be addressed.
It is in fact convenient to restrict the accessible CV space such that the region in which the system is totally converted into liquid is not explored.
This region is not of interest since here we focus on the nucleation barrier and restricting the CV space accelerates the convergence.
We shall use $p(s)$ to limit the exploration of the CV space.
In particular we shall choose a $p(s)$ that is uniform until a value $s_0$ and vanishes smoothly beyond it, i.e.:
\begin{equation}
	p(s)=
	\begin{cases}
		\frac{1}{C} \: &  \mathrm{if} \: s<s_0 \\
		\frac{1}{C} e^{-\frac{1}{2} \beta \kappa (s-s_0)^2 } \: & \mathrm{if} \: s>s_0\\
	\end{cases} ,
\end{equation}
where $s_0$ should lie beyond the barrier region, $\kappa$ is a constant that determines how fast $p(s)$ goes to zero, and $C$ is a normalization constant.
The bias potential would be able to produce this $p(s)$ provided that it had sufficient variational flexibility.
Since we use instead a bias potential with minimal flexibility, $V(s)$ must be constructed in such a way that it is capable of satisfying equation \eqref{optimal_bias} in all the CV space.
A bias potential that is able to do so is:
\begin{equation}
	V(s)=
	\begin{cases}
		-\Delta F^{CNT}(s;\Delta\mu,\sigma) \: &  \mathrm{if} \: s<s_0 \\
		-\Delta F^{CNT}(s;\Delta\mu,\sigma) + \frac{1}{2}\kappa (s-s_0)^2 \: & \mathrm{if} \: s>s_0\\
	\end{cases}
  .
  \label{eq:bias_with_barrier}
\end{equation}
Appropriate values for $\kappa$ and $s_0$ can be easily chosen based on a very approximate knowledge of the free energy landscape.
In Figure \ref{fig:methodIllustration} the functional forms of $F(s)$, $V(s)$ and $p(s)$ are depicted.

\subsection{Optimization algorithm}

Inserting \eqref{eq:bias_with_barrier} into $\Omega[V]$, the functional becomes a function of $\Delta\mu$ and $\sigma$, $\Omega(\Delta\mu$,$\sigma)$.
In order to optimize the functional we need to evaluate the gradient: 
\begin{align}
    \frac{\partial\Omega(\Delta\mu,\sigma)}{\partial\:\Delta\mu}= & \langle s \rangle_V - \langle s \rangle_p  \nonumber \\
    \frac{\partial\Omega(\Delta\mu,\sigma)}{\partial\sigma}= & - \langle s^{2/3} \rangle_V + \langle s^{2/3} \rangle_p
\end{align}
and the Hessian matrix:
\begin{equation}
H(\Delta\mu,\sigma)= 
  \begin{bmatrix}
    H_{\Delta\mu\:\Delta\mu} & H_{\Delta\mu\:\sigma}  \\
   H_{\sigma\:\Delta\mu}  & H_{\sigma\:\sigma} \\
  \end{bmatrix}
= \beta \cdot
  \begin{bmatrix}
    \langle s^2 \rangle_V - \langle s \rangle_V^2     &
    -\langle s^{5/3} \rangle_V+\langle s \rangle_V \langle s^{2/3} \rangle_V \\
    -\langle s^{5/3} \rangle_V+\langle s \rangle_V \langle s^{2/3} \rangle_V &
    \langle s^{4/3} \rangle_V-\langle s^{2/3} \rangle_V^2\\
  \end{bmatrix} .
\end{equation}
These expressions involve only expectation values either in the biased ensemble $\langle \: \rangle_V$ or over the target distribution $\langle \: \rangle_p$.

Crucial to a succesful optimization is the use of the averaged stochastic gradient descent algorithm \cite{Bach13}.
In the present case this algorithm can be written as an iterative procedure with a fixed step size $\mu$:
\begin{align}
  \Delta\mu^{(k + 1)}  & = \Delta\mu^{(k)} - \mu
  \left [
  \frac{\partial\Omega}{\partial\:\Delta\mu}
  + H_{\Delta\mu\:\Delta\mu} \cdot \left (\Delta\mu^{(k)}-\overline{\Delta\mu}^{(k)} \right)
  + H_{\Delta\mu\:\sigma} \cdot \left (\sigma^{(k)}-\overline{\sigma}^{(k)} \right)
  \right] \nonumber \\
  \sigma^{(k + 1)}  & = \sigma^{(k)} - \mu
  \left [
  \frac{\partial\Omega}{\partial\sigma}
  + H_{\sigma\:\Delta\mu} \cdot \left (\Delta\mu^{(k)}-\overline{\Delta\mu}^{(k)} \right)
  + H_{\sigma\:\sigma} \cdot \left (\sigma^{(k)}-\overline{\sigma}^{(k)} \right)
  \right],
\end{align}
where $\Delta\mu$ and $\sigma$ are the instantaneous parameters whereas $\overline{\Delta\mu}^{(k)} = k^{-1} \sum_{i=1}^k \Delta\mu^{(i)}$ and $\overline{\sigma}^{(k)} = k^{-1} \sum_{i=1}^k \sigma^{(i)} $ are their averaged counterparts.
At each iteration $k$, both the gradient and the Hessian matrix are estimated in the bias ensemble with a bias potential given by the averaged parameters $\overline{\Delta\mu}^{(k)}$ and $\overline{\sigma}^{(k)}$.

In previous calculations we have used only the diagonal part of the Hessian.
Here the use of the full Hessian is essential for a succesful optimization.
In fact the minimization problem is ill-conditioned with a condition number $\sim 10^4$.

Although using the full Hessian allowed us to reach the minimum, we found that a faster convergence could be achieved by making the following change of variables,
\begin{equation}
s(x)=N_0 \left (\frac{x+1}{2} \right )^3,
\end{equation}
where $N_0$ is a number slightly larger than the point $s_0$ where $p(s)$ starts decaying towards zero.
This change of variables is akin to the one described in section \ref{sec:ClassicalNucleationTheory} that transforms the number of atoms $n$ in the droplet into its radius $R$.
Therefore $x$ is related to a characteristic length of the droplet.
With this change of variables the new CV is defined in the interval $[-1,1]$ and, for $s<s_0$, $V(x;\Delta\mu,\sigma)$ can be written as a polynomial:
\begin{equation}
V(x;\Delta\mu,\sigma)= N_0 \Delta\mu \left (\frac{x+1}{2} \right)^3 - N_0^{2/3} \sigma \left ( \frac{x+1}{2} \right)^2 .
\label{eq:bias_change_variables}
\end{equation}
It is therefore natural to express $V(x;\Delta\mu,\sigma)$ in terms of Chebyshev polynomials,
\begin{equation}
   V(x;\Delta\mu,\sigma)=\sum_{i=0}^3 \alpha_i \cdot T_i(x)
\end{equation}
where $T_i(x)$ is the Chebyshev polynomial of degree $i$.
Comparison with the expression in equation \eqref{eq:bias_change_variables} gives a set of relations between the $\alpha_i$, and $\Delta\mu$ and $\sigma$: 
\begin{align}
        \alpha_0 = & \frac{5 \Delta\mu \: N_0}{16}  - \frac{3 \sigma \: N_0^{2/3}}{8}  \nonumber \\
        \alpha_1 = & \frac{15 \Delta\mu \: N_0}{32}  - \frac{\sigma \: N_0^{2/3}}{2} \nonumber \\%4 \alpha_2 - 9 \alpha_3 \nonumber \\
        \alpha_2 = & \frac{3 \Delta\mu \: N_0}{16}  - \frac{\sigma \: N_0^{2/3}}{8} \nonumber \\
        \alpha_3 = & \frac{\Delta\mu \: N_0}{32} .
	\label{eq:constraints}
\end{align}
We can now use, say, $\alpha_2$ and $\alpha_3$ as variational parameters, constraining the other two via equations \eqref{eq:constraints}.
In these new variables, the problem is better behaved and the condition number of the Hessian is reduced to the manageable value of $\sim 5$.
Although $V(s;\alpha_2,\alpha_3)$ depends only on two variables, the fact that it can be formally expressed as a linear expansion in orthogonal polynomials allows us to use the optimization machinery previously developed to handle this case \cite{Valsson14}.

%%%%%%%%%%%%%%%%%%%%%%%%%%%%%%%%%%%%%%%%%%%%%%%%%%%%%%%%%%%%%%%%%%%%%%%%%%%%%%%
\section{Computational details}
%%%%%%%%%%%%%%%%%%%%%%%%%%%%%%%%%%%%%%%%%%%%%%%%%%%%%%%%%%%%%%%%%%%%%%%%%%%%%%%

We have studied condensation from the vapor phase in a Lennard-Jones system in which the interaction potential was truncated and shifted at a cutoff radius $r=2.5 \sigma_{LJ}$ , with $\sigma_{LJ}$ the particle diameter.
All MD simulations were performed using LAMMPS \cite{Plimpton95} patched with a private development version of the PLUMED 2 enhanced sampling plug-in \cite{Tribello14}.
In the following we shall measure all quantities in Lennard-Jones units \cite{FrenkelBook}, such that the Lennard-Jones well depth $\epsilon$ is the unit of energy and the Lennard-Jones diameter $\sigma_{LJ}$ is the unit of length.
In the definition of the liquid-like atoms we used the values $r_c=1.5$ and $n_c=5$ as suggested in ref.\ \citenum{tenWolde98}.

Periodic boundary conditions and a time-step of 0.001 were used in the simulations \cite{FrenkelBook}.
In all cases the stochastic velocity rescaling thermostat \cite{Bussi07} and the isotropic version of the Parrinello-Rahman barostat \cite{Parrinello81} were employed.
The relaxation time for the thermostat and the barostat were 0.05 and 50, respectively. 
We employed cubic boxes with different number of particles and a temperature of 0.741 ($T_c=1.085$).
The target pressure of the barostat was set to 0.016 .
This system setup is similar to that of ref.\ \citenum{tenWolde98}, although the supersaturation is higher in our case.

Each iteration in the optimization of $\Omega$ corresponded to 500 MD steps and the step size $\mu$ in the optimization was chosen to be $0.001$.
In all cases 4 multiple walkers were employed, starting half of them in the vapor basin and the rest beyond the nucleation barrier.
Due to the highly non-local nature of the basis sets employed, the use of multiple walkers proved to be instrumental in accelerating the optimization.
The initial variational parameters were taken as $\Delta\mu=0$ and $\sigma=0$ such that initially the bias was $V(s)=-\frac{1}{\beta}\log p(s)$.
The parameters of $p(s)$ were $s_0=120$ and $\kappa=0.1$.

%%%%%%%%%%%%%%%%%%%%%%%%%%%%%%%%%%%%%%%%%%%%%%%%%%%%%%%%%%%%%%%%%%%%%%%%%%%%%%%
\section{Results}
%%%%%%%%%%%%%%%%%%%%%%%%%%%%%%%%%%%%%%%%%%%%%%%%%%%%%%%%%%%%%%%%%%%%%%%%%%%%%%%

\begin{figure}[t]
\centering \includegraphics[width=0.95\textwidth]{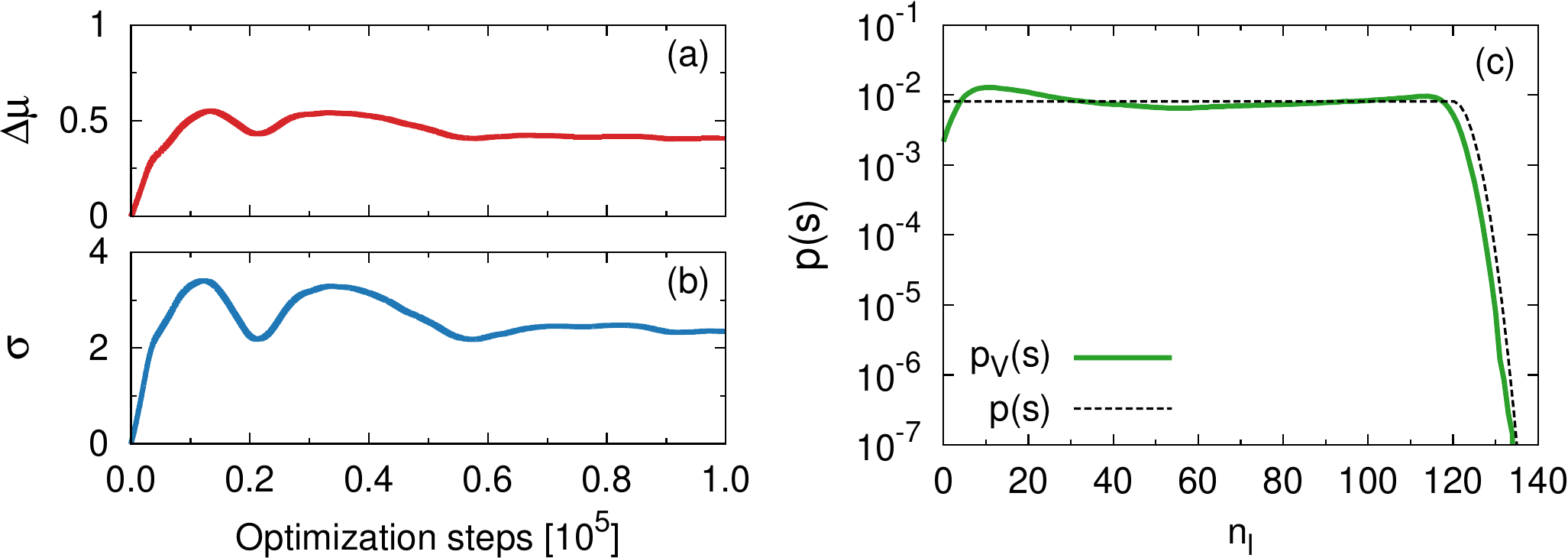}
\caption{\label{fig:convergence_and_p}
(a)-(b) Evolution of $\Delta\mu$ and $\sigma$ during the optimization process for a system of 216 particles.
(c) Comparison between the sampled probability distribution of the CV $n_l$ $p_V(s)$ and the target probability distribution $p(s)$.
}
\end{figure}

As an example of a typical behavior of the optimization process, we show in Figure \ref{fig:convergence_and_p} (a)-(b) the convergence of the variational parameters $\Delta\mu$ and $\sigma$ as a function of the number of optimization steps in a system with 216 atoms.
If we use Lennard-Jones parameters appropriate to argon, the total length of the optimization corresponds to a $\sim 100$ ns long simulation.
A measure of the quality of the variational ansatz in equation \eqref{eq:bias_model} is how much the biased distribution $p_V(s)$ differs from the target distribution $p(s)$ (see Figure \ref{fig:convergence_and_p}(c)).
According to equation \eqref{p_of_s_biased_ensemble} these should be identical.
Indeed, they are very close but not totally identical.
The small discrepancies are a result of the limited variational flexibility of $V(s)$ (equation \eqref{eq:bias_model}).

\begin{figure}[t]
\centering \includegraphics[width=0.95\textwidth]{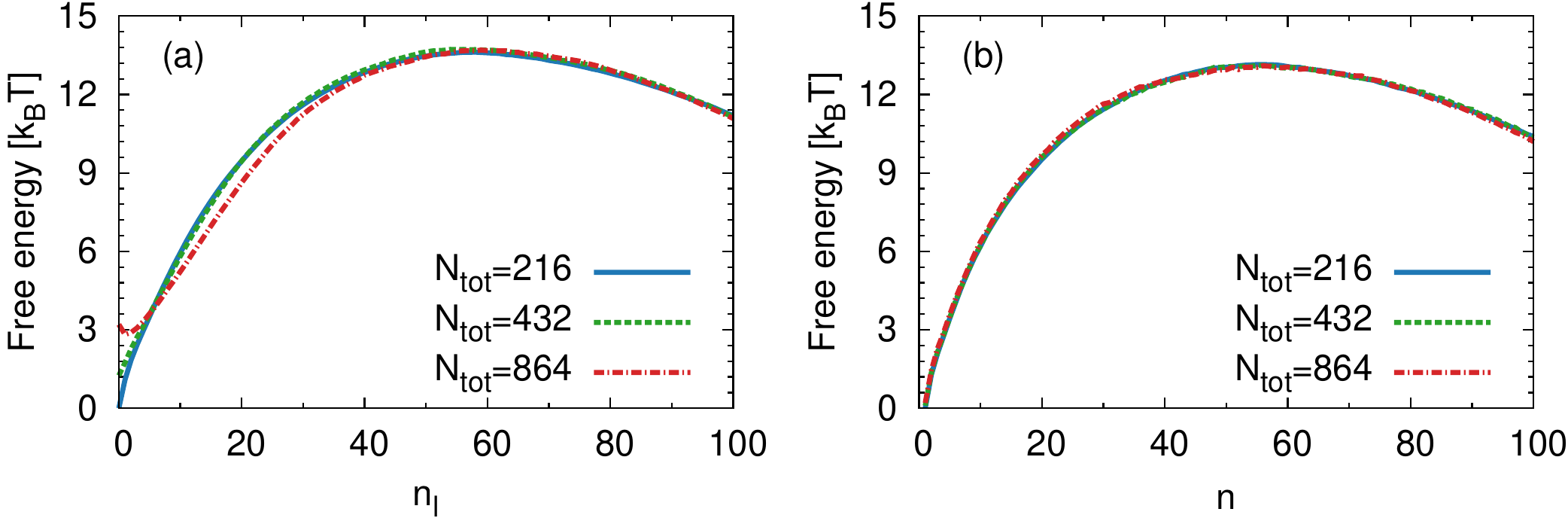}
\caption{\label{fig:compareCV_combine} 
 (a) Reweighted free energy as a function of $n_l$ for system sizes $N_{tot}=216,432$ and $864$. For small $n_l$ the curves show discrepancies.
 (b) Reweighted free energy associated to $n$ for the same system sizes as above. All free energies are equal within the statistical error.
}
\end{figure}

We now turn our attention to the behavior of the free energy as a function of the collective variable $n_l$ for three different system sizes (see Figure \ref{fig:compareCV_combine}(a)).
By and large, the curves are rather similar, however, some differences are observed.
This is not surprising since the probability of observing several liquid clusters at one time depends on the system size.
This is particularly evident for the low $n_l$ region on which the largest size dependence is observed.
This behavior has already been reported for the free energy associated with the size of the largest cluster \cite{Maibaum08comment}.
For larger $n_l$ the probability of finding more than one cluster is small, at least for the system sizes studied here, and the size effects are negligible.
These finite size effects are quantified in Table \ref{tbl:fit_cnt} where we compare the estimates of $\Delta\mu$ and $\sigma$ obtained from minimizing $\Omega$ to a fit to $F(n_l)$ of the CNT expression.
It is seen that the smaller the system, the smaller the deviation of $F(n_l)$ from a CNT-like behavior.
However if we reweight our data so as to obtain $F(n)$ (see Figure \ref{fig:compareCV_combine}(b)), all the finite size effects dissapear.
Details of the reweighting procedure and of the calculation of $F(n)$ can be found in the SI.
The results can be all fitted to the CNT expression and give what is possibly our best estimate for these parameters (see Table \ref{tbl:fit_cnt}), assuming that our data can be described by CNT.
The values obtained are consistent with the estimates of $\Delta\mu=0.530$ and $\sigma=2.85$ that can be calculated using a coexistence pressure $P_0=0.00783$ obtained from Gibbs ensemble simulation \cite{tenWolde98} and a surface free energy $\gamma=0.494$ \cite{tenWolde98,Holcomb93}.

\begin{table}[h]
\small
  \caption{
We compare estimations of $\Delta\mu$ and $\sigma$ for different system sizes. 
From left to right we show the direct result of the optimization at $5 \cdot 10^5$ steps, a fit of the CNT expression to $F(n_l)$ in the interval [0,N], and a similar fit to $F(n)$ in the interval [1,N].
In the last two cases, we show also the error associated to the fit.
}
  \label{tbl:fit_cnt}
  \begin{tabular*}{1.0\textwidth}{@{\extracolsep{\fill}}ccccccccc}
    \hline
                        & \multicolumn{2}{c}{Optimization}  & \multicolumn{3}{c}{Fit $F(n_l)$} & \multicolumn{3}{c}{Fit $F(n)$} \\ 
\cline{2-3} 
\cline{4-6} 
\cline{7-9} 
    System size & $\Delta\mu$ & $\sigma$ & $\Delta\mu$ & $\sigma$ & rms error & $\Delta\mu$ & $\sigma$ & rms error \\
    \hline
    216 & 0.42(1) & 2.4(1) & 0.409(6) & 2.38(3) & 0.39  & 0.439(2) & 2.51(1) & 0.12 \\
    432 & 0.41(2) & 2.4(3) & 0.408(8) & 2.37(4) & 0.54  & 0.432(2) & 2.48(1) & 0.12 \\
    864 & 0.37(4) & 2.2(5) & 0.38(1)  & 2.22(7) & 0.83  & 0.436(2) & 2.49(1) & 0.11 \\
    \hline
  \end{tabular*}
\end{table}

From $F(n)$ we can estimate the nucleus size and the barrier height.
The nucleus size $n^*$ at the supersaturation condition studied here corresponds to 56 atoms.
We estimated the barrier height defined as $F(n^*)$-$F(1)$ to be  $\sim 13$ $k_{B}T$.

In order to test the correctness of the approach described in this work, we have performed a benchmark simulation employing well-tempered metadynamics\cite{Barducci08}.
The details of this simulation can be found in the SI.
The free energies obtained with the methodology described in this work are equal to those calculated employing well-tempered metadynamics within the statistical error (see Figure S1 in the SI).

%%%%%%%%%%%%%%%%%%%%%%%%%%%%%%%%%%%%%%%%%%%%%%%%%%%%%%%%%%%%%%%%%%%%%%%%%%%%%%%
\section{Discussion and conclusions}
%%%%%%%%%%%%%%%%%%%%%%%%%%%%%%%%%%%%%%%%%%%%%%%%%%%%%%%%%%%%%%%%%%%%%%%%%%%%%%%

We have developed an enhanced sampling method based on the introduction of a bias potential with a physically motivated functional form.
In particular, in the context of nucleation, we have employed the functional form of classical nucleation theory for the free energy of formation of a cluster.
This idea is put into practice employing a variational principle that allows the estimation of the parameters of the model.
In this way the bias potential compensates the underlying free energy of the system.

Our results are encouraging, however they underline the fact that attention must be paid to the choice of the collective variables.
In particular for the system of interest here, the choice of $n_l$ as CV has clear computational advantages but it is best applied to small systems where $F(n_l)$ is very close to the free energy associated to the cluster size distribution $F(n)$, that is system size independent.
It must also be noted that for small systems the values for $\Delta\mu$ and $\sigma$ obtained via the optimization, are very close to those obtained by reweighting the trajectory to get the cluster size distribution and fitting the CNT expression to the results.
This might provide and expedite way to estimate the vapor pressure and the surface energy, two quantities of great practical interest.

The lesson learned here will be applied to nucleation rates calculation.
In fact, given the results obtained, one could use $n_l$ as the collective variable and employ the approaches suggested either in ref.\ \citenum{Tiwary13} or ref.\ \citenum{McCarty15}.
Both methodologies rely on introducing a bias potential that leaves the transition region between metastable states untouched.
Under this assumption the physical transition time ($\tau$) can be related to the one calculated in a biased simulation ($\tau_V$) by\cite{Tiwary13,Voter97,Grubmuller95},
\begin{equation}
	\tau = \tau_V \: \langle e^{\beta V(s,t)} \rangle_V .
	\label{eq:rates}
\end{equation}
However, the two methodologies differ from each other in the manner in which the bias potential is constructed.
On the one hand, ref.\ \citenum{Tiwary13} describes a metadynamics based methodology with infrequent deposition of kernels.
If the transition is rare but fast then the procedure leads to bias free transition regions therefore fulfilling the assumption that leads to equation \eqref{eq:rates}.
On the other hand, the approach described in ref.\ \citenum{McCarty15} is based on the construction of a bias potential by means of the variational principle\cite{Valsson14} also used in the present work.
This bias potential floods the free energy surface up to a predefined energy level.
By construction this approach guarantees bias free transition states and accurate times can be extracted using equation \eqref{eq:rates}.
We point out that the independence of $F(n)$ from the system size provides a strong encouragement to limit ourselves to the study of small systems.

Finally, our results provide yet another confirmation of the validity of CNT for liquid-vapor nucleation.

%%%%%%%%%%%%%%%%%%%%%%%%%%%%%%%%%%%%%%%%%%%%%%%%%%%%%%%%%%%%%%%%%%%%%
\section*{Acknowledgements}
%%%%%%%%%%%%%%%%%%%%%%%%%%%%%%%%%%%%%%%%%%%%%%%%%%%%%%%%%%%%%%%%%%%%%

P.M.P would like to thank Matteo Salvalaglio and Federico Giberti for useful discussions.
The authors acknowledge funding from the National Center for Computational Design and Discovery of Novel Materials MARVEL and the European Union Grant No. ERC-2014-AdG-670227 / VARMET. 
The computational time for this work was provided by the Swiss National Supercomputing Center (CSCS).

%\bibliography{mybib}
%\bibliographystyle{unsrt}

\end{document}